# Two are not always better than one:

# Role specialization is an important determinant of

# collaborative task performance

Asuka Takai,[#*1†] Qiushi Fu,[#2‡] Yuzuru Doibata,[1] Giuseppe Lisi,[1%]
Toshiki Tsuchiya,[2§] Keivan Mojtahedi,[2!] Toshinori Yoshioka,[1,3]
Mitsuo Kawato,[1] Jun Morimoto,[*1] Marco Santello[*2]

[1]Brain Information Communication Research Laboratory Group,
Advanced Telecommunications Research Institute International,
2-2-2 Hikaridai Seika-cho, Soraku-gun, Kyoto, 619-0288 Japan

[2]School of Biological and Health Systems Engineering,
Arizona State University,
Tempe, AZ 85287 USA

[3]XNef,
2-2-2 Hikaridai, Seika-cho, Soraku-gun, Kyoto 619-0288, Japan

[#]Equal contribution

[*]Co-corresponding authors
Email: atakai@atr.jp; xmorimo@atr.jp; marco.santello@asu.edu

Current address:
  [†] Graduate School of Engineering Division of Mechanical Engineering, Osaka Metropolitan University,
    3-3-138 Sugimoto, Sumiyoshi-ku, Osaka-shi, Osaka 558-8585 Japan
  [‡] Mechanical and Aerospace Engineering, University of Central Florida,
    Orlando, FL 32816
  [%] Scuderia Alpha Tauri F1 Team,
    Bicester, Oxfordshire OX26 4LD, GB
  [§] SAP Japan Co., Ltd.,
    Mitsui Bussan Building 1-2-1, Otemachi Chiyoda-ku, Tokyo 100-0004, Japan
  [!] Dexcom, Inc.,
    6340 Sequence Dr., San Diego, CA 92121

**ABSTRACT**

Collaboration frequently yields better results in decision making, learning, and haptic interactions than when these actions are performed individually. However, is collaboration always superior to solo actions, or do its benefits depend on whether collaborating individuals have different or the same roles? To answer this question, we asked human subjects to perform virtual-reality collaborative and individual beam transportation tasks. These tasks were simulated in real-time by coupling the motion of a pair of hand-held robotic manipulanda to the virtual beam using virtual spring-dampers. For the task to be considered successful, participants had to complete it within temporal and spatial constraints (1< and <2 s and <1.5° maximum tilt, respectively). While the visual feedback remained the same, the underlying dynamics of the beam were altered to create two distinctive task contexts which were determined by a moving pivot constraint. When the pivot was placed at the center of the beam, two hands contribute to the task with symmetric mechanical leverage (symmetric context). When the pivot was placed at the left side of the beam, two hands contribute to the task with asymmetric mechanical leverage (asymmetric context). Participants performed these task contexts either individually with both hands (solo), or collaboratively by pairing one hand with another one (dyads). We found that dyads in the asymmetric task context performed better than solos. In contrast, solos performed the symmetric task context better than dyads. Importantly, we found that two hands took different roles in the asymmetric context for both solos and dyads, with the hand on the side closer to the pivot leading the movement and exerting more force. In contrast, the contribution from each hand was statistically indistinguishable in the symmetric context. Our findings suggest that better performance in dyads than solos is not a general phenomenon, but rather that collaboration yields better performance only when role specialization emerges in dyadic interactions.

# I. INTRODUCTION

It is generally assumed that collaboration can lead to better performance than efforts made by individuals. This phenomenon, also known as the "assembly bonus effect"[1], in which group performance exceeds the performance of individual group members, has been reported in various types of collaboration, e.g., decision-making tasks[2], language acquisition[3], cards sorting tasks[4], and programming tasks[5].

In the domain of haptic collaboration, i.e., joint actions performed by physically-interacting individuals, investigations on whether collaborating agents can outperform individual agents has yielded mixed results[6]. Some studies demonstrated that paired individuals perform better than single agents. Reed and colleagues[7] found dyads can perform a target acquisition task with a rotational crank faster than individuals. Ganesh and colleagues[8], using a two-dimensional continuous target tracking task, found that dyadic performance was characterized by smaller target tracking error than solo performance. However, other studies did not find performance advantages in dyads over solos. Van der Wel and colleagues[9] reported that dyads performed a pendulum swing task at the same level as solos. In tasks that require keeping an object stable against gravity, several studies have found that dyadic performance was, on average, similar to solos[10, 11, 12].

There are several potential factors that may account for the differences in these findings.

One factor could be the number of effectors involved in the task. Dyads tend to outperform solos if the task can be performed unimanually by two individuals[7, 8, 13, 14]. This can be explained by the biomechanical advantages associated with the involvement of more effectors during dyadic cooperation, i.e., two arms from two agents versus one arm from one agent. Another factor could be the extent to which paired individuals can focus on the control of different aspects of the task, i.e., role specialization[6]. This factor could be particularly important when two effectors are required to perform a given task. For instance, van Oosterhout and colleagues[13] assigned two participating arms asymmetrical and independent roles: one arm controlled the vertical motion of the object whereas the other arm controlled the horizontal motion in a tele-operation task. This study demonstrated superior performance by dyads over bimanual solo when roles are pre-assigned. However, it remains unknown whether role specialization that emerges spontaneously during collaboration may benefit dyadic more than solo performance. Spontaneous role specialization has been observed in different physically-coupled motor collaboration tasks[15, 16, 17, 18, 19], but these studies did not examine solo performance of bimanual tasks.

To address these gaps, we designed an experimental study of a virtual beam transport tasks that must be completed by two effectors belonging to one agent (bimanual condition) or two agents (dyadic condition) to minimize the beam's tilt. Importantly, the dynamics of the virtual beam can be altered to give the two participating effectors symmetric or asymmetric implicit mechanical leverages. One of our recent studies has shown that, when asymmetric mechanical leverages were given, stable and consistent role specialization can emerge spontaneously and gradually improves the performance of the task in dyads[19]. Furthermore, another recent study of ours demonstrated that no consistent role specialization was observed when symmetric mechanical leverages were given to dyads[20]. Therefore, we hypothesized that spontaneous role specialization is a necessary condition for superior performance, which can only be found in the asymmetric but not the symmetric context of our task.

## II. METHODS

### A. Participants

Sixteen healthy participants provided written informed consent before participating to the experiments. Their handedness was tested using the Edinburgh Handedness Inventory, confirming that all were right-handed. The ATR Review Board Ethics Committee approved the study protocol.

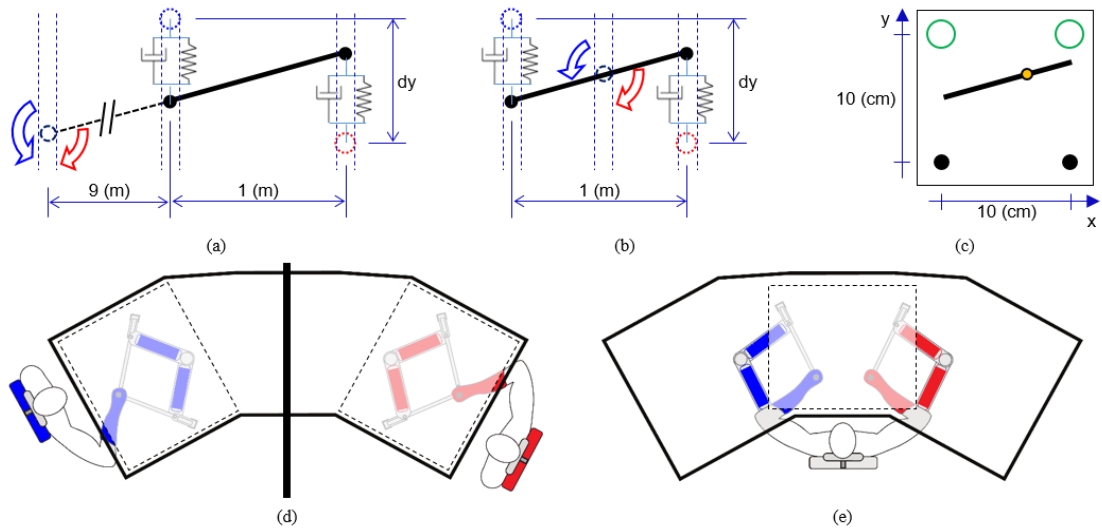

**Figure 1**: **Task, manipulanda and experimental contexts. (a)**. Task dynamics of the left-pivoted beam (asymmetric context). The virtual beam, shown as a solid black line, behaves as if its rotational point (pivot, dashed circle) is located at the left end. Participants' hands (dashed red and blue dotted circles) are connected to the beam ends (black dots) via simulated spring-dampers. The distance between each handle was 1 m. The distance from the pivot to the left handle was 9 m. We defined this experimental context as "asymmetric" based on the unequal distance between each handle and the pivot. **(b).** Task dynamics of the center-pivoted beam (symmetric context). This virtual beam behaves as if its pivot is located at the center (the pivot-to-hand distance for both hands is the same), and therefore we defined this experimental context as "symmetric." **(c).** Visual feedback of the task. The visualized beam is shown as a solid black line. Participants transported the beam from the start position (black dots) to the target location (green circles). The transport distance was

10 cm on the display. Beam tilt was rendered as a yellow dot sliding along the beam (sensitivity: 0.4 cm/degree). **(d).** Dyadic condition. Each participant moved a robotic manipulandum located under a white and opaque table so that the participant could not see her/his own hands. Positions of the two handles and force applied to them controlled the movement of the virtual beam. Simulated physical interaction force provided haptic feedback to participants through the manipulanda. Visual feedback was projected on the table by projectors. Participants wore headphones to eliminate aural cues while blackboards placed between them prevented visual cues. Participants were informed of the task goal (the maximum absolute beam tilt and the range of transportation time) and were told that the partner was performing a different task. **(e).** Solo condition. Individual participants moved both manipulanda. Solo participants received the same visual feedback provided to dyads. The actual hand positions were 7.5 cm from the ends of the visualized beam in the x-direction (the physical distance between the hands was 25 cm). Participants wore headphones to match conditions in the dyad experiments. Solo participants were informed of the task goal as done with participants to the dyadic condition.

B. *Experimental setup*

We implemented a virtual beam transportation task which was controlled by moving a pair of robotic manipulanda (TVINS)[8, 21, 22]. These robots are composed of parallel links driven by electric motors. The handles of the robots were aero-magnetically floated on the table to reduce friction. Forces exerted by participants holding the handle(s) were measured at each handle with a six-axis force sensor. The handle forces and positions were collected at a sampling rate of 2 kHz to control the virtual beam movement. The spatial configuration of the robots was based on the number of performers. In the paired condition, both participants grasped the robot handle with their right hands, and the robots and the seating of the participants were separated by a wall such that they did not see each other (Fig. 1d).

In the bimanual solo condition, the robots were placed closely to allow participants to grasp robot handles with both hands (Fig. 1e). For both conditions, we set the shoulder-elbow-wrist in the horizontal plane with the handle in the same plane, which is beneath an opaque display surface of the task visual feedback. Participants' forearms were out of view and rested on a cuff that was supported horizontally. Moreover, the robots were programmed to only move in one direction (forward or backward with respect to the participants; y-direction, Fig. 1c).

Visual feedback of the task was the same in all experimental conditions and contexts. Specifically, each participating hand must move one end of a virtual beam (10-cm long) from a starting position toward a target position (10-cm away). They were allowed to overshoot the target, but they were asked to keep the beam balanced within a maximum absolute tilt of 1.15° during beam transport. Participants were required to transport the beam in 1 to 2 seconds.

Importantly, the virtual beam visual movement can be governed by two different dynamics: one with a hidden virtual pivot at the left end and the other with a hidden pivot at the center (Fig. 1a and 1b, respectively). In both of these task contexts, robotic handles were coupled to the ends of the virtual beam with virtual spring-dampers, such that a virtual force was generated based on the relative movement between the handle and the beam. The underlying dynamics of these two contexts differed depending on the location of the pivot.

For the asymmetric context, the pivot was located to the left side of the beam and two hands acted on the beam through asymmetric moment arms (length ratio 9:10 for left versus right). Moreover, as both hands are on the right side of the pivot, they produce counterclockwise (CCW) moment if they are moving in front of the beam, and clockwise (CW) moment if they are moving behind the beam. In contrast, for the symmetric context, the pivot was located at the center of the beam and two hands acted on the beam through symmetric moment arms (same length). Furthermore, the hands on the left and right side of the beam have opposite moment production capability.

Beam linear movement was simulated using the equation of motion (Eq. (1)). Beam rotary motion was simulated using Eq. (2) for the left-pivoted beam model and Eq. (3) for the center-pivoted beam model:

$$\widetilde{F_{1y}} + \widetilde{F_{2y}} = M\ddot{y}_p + B\dot{y}_p \qquad (1)$$

$$L_2\widetilde{F_{2y}} + L_1\widetilde{F_{1y}} = I\ddot{\theta} + B_p\dot{\theta} \qquad (2)$$

$$L/2(\widetilde{F_{2y}} - \widetilde{F_{1y}}) = I\ddot{\theta} + B_p\dot{\theta} \qquad (3)$$

Parameters of the beam and virtual spring and damper are shown in Table 1. In addition to the participant interaction force measured at handles, the input force for the model was derived from deviations between the handle and the beam in position and velocity:

$$\widetilde{F_{ny}} = K_H(y_n - y_{n,beam}) + B_H(\dot{y}_n - \dot{y}_{n,beam}) + G \cdot F_{ny} \qquad n = 1,2 \qquad (4)$$

The following equation calculated the feedback force from the beam:

$$F_{ny,fb} = -K_T(y_n - y_{n,beam}) \quad n = 1,2 \quad (5)$$

Table 1. Parameters of the virtual beam

| Coefficients | Value |
|---|---|
| $M$ Weight of the virtual beam (kg) | 1 |
| $L$ Length of the virtual beam (m) | 1 (Center-pivoted) <br> 10 (Left-pivoted) |
| $B$ Virtual beam transportation viscosity (Ns/m) | 0.1 |
| $B_p$ Virtual beam rotational viscosity (Nms) | 0.1 |
| $K_H$ Force input spring stiffness (N/m) | 100 |
| $B_H$ Force input damper viscosity (Ns/m) | 5 |
| $G$ Force input gain (-) | 0.1 |
| $K_T$ Force feedback spring stiffness (N/m) | 400 |

C. *Protocol*

We investigated the effects of two factors on task performance: Condition (the number of agents: dyads versus solos) and Context (asymmetric versus symmetric beam dynamics). The order of presentation of the experimental condition was counterbalanced such that half of the participants executed the solo task first and then moved to the paired task, and vice versa for the other half of the participants. Within each condition, the symmetric condition was always performed before the asymmetric condition. Each participant started to move the manipulandum when the start color on the visual display changed. On each trial, the participant then moved the beam only in one direction until the beam reached the target, at which point the robotic manipulandum returned the arm to the initial position. Participants re-entered the start area by themselves. In all conditions and contexts, the experiment continued until participants performed 40 successful trials or a maximum of 250 total trials.

D. *Quantification of performance*

The task performance of dyads and individuals was quantified by three metrics: number of trials performed to achieve 40 successful trials, beam transport time and maximum absolute beam tilt angle. The number of total trials quantifies the overall task performance, with a larger number indicating a greater difficulty with complying with the task requirements. Beam transport time is defined as the time between the start of the trial to when both ends of the displayed beam passed the target, and therefore quantifies task completion speed. Maximum absolute beam tilt angle, measured during beam transport, quantifies the accuracy of task performance. Note that the task did not explicitly require participants to minimize beam tilt. However, a smaller maximum absolute beam tilt angle would denote subjects adopting a larger 'safety margin' with respect to the required task success threshold.

For both the maximum beam tilt angle and transport time, we computed the mean value of the first and last five successful trials to represent the task performance at the beginning and end of the block of trials for each condition, respectively. This enabled us to measure the extent to which performance may have improved across successful trials[19]. Furthermore, to quantify the extent to which task performance might have differed between solo and dyadic conditions, we computed the difference in these three performance metrics between these two conditions.

For statistical analysis, we excluded one pair of participants because their behavior was identified as an outlier (maximum beam tilt angle was greater than $\pm 2.5$ standard deviations). Subsequently, we used one-sample t-tests for the dyad-solo difference for the total number of successful trials to quantify whether there were differences between two conditions. We used two-way repeated ANOVA (two factors; Context: symmetric, asymmetric; Trial: first and last five trials) to assess differences between dyadic and solo conditions in maximum beam tilt angle and beam transport time. Statistically significant interactions were further analyzed using post-hoc comparisons, and one-sample t-tests were used for each condition to identify the presence of non-zero differences between dyadic and solo conditions. Bonferroni corrections were applied to comparisons within each metric.

E. *Quantitative identification of participants' roles*

Prior research of physical joint actions has revealed that agents can take on different roles, e.g., "active-passive" and "acceleration-deceleration"[15], "executor-conductor"[23], and "pushing-pulling"[18]. In the present work, both participants in dyads or both hands of individual agents must coordinate their pushing/pulling actions to achieve the task goal, but two hands may not move at the same speed and therefore one could be spatially and/or temporally leading the other. To quantify this phenomenon, we will use the same definitions developed by previous studies of joint motor tasks: the "leader" leads the actions while the action by the "follower" lags[24]. Thus, we compared the mean position difference of left versus right handles within each trial. A positive value represents that the left agent/hand spatially led the right one. Other joint motor task studies have shown that cooperating participants often do not make equal contributions to the task[25]. Therefore, we also compared the difference in mean absolute force between the left and right hands within each trial. A positive value represents that the left agent/hand used more force than the right one.

To assess the effect of repeated practice across trials, we computed the average value of the above two role-specialization metrics (handle position difference and force difference) from the first and last five successful trials to evaluate the relative contribution of the two hands at the initial and final portion of a block of trials for each condition, respectively. For statistical analysis, we also excluded the same pair of participants identified from the performance analysis. Subsequently, we used two-way repeated ANOVA (two factors; Context: symmetric, asymmetric; Trial: first and last five trials) for both metrics. Statistically significant interactions were further analyzed using post-hoc comparisons, and one-sample t-tests were used for each condition to determine the presence of non-zero differences between the left and right agents/hands. Bonferroni corrections were applied to comparisons within each metric.

## III. RESULTS

Dyads and solos were able to accomplish 40 successful trials in all conditions and contexts. This section describes how participants executed our task in the symmetric and asymmetric contexts and whether task execution differed when comparing collaborative versus individual performances in each context.

A. ***Dyads performed better than solos in the asymmetric, but not in the symmetric context***

To be successful, all participants had to transport the virtual beam by ensuring a tilt angle less than 1.15°. Dyads tilted the beam less than solos in the asymmetric context but tilted the beam more than solos in the symmetric context (main effect of Context: $F(1,13) = 104.5$, $p < 0.001$, Fig. 2a). Furthermore, for the asymmetric conditions, one-sample t-tests revealed that dyads exhibited significantly smaller tilt than solos in the last five successful trials (p = 0.001), but not in the first five successful trials. In contrast, dyads exhibited significantly larger tilt than solos in the symmetric conditions for both first and last five successful trials (p < 0.001 and p < 0.01, respectively).

We also asked all participants to execute the task in no less than 1 second and no more than two seconds for performance to be considered successful. We found a significant Context × Trial interaction (F(1,13) = 6.33, p = 0.026, Fig. 2b). However, post-hoc comparisons did not reveal significant differences between conditions after Bonferroni corrections. One-sample t-test showed that transport time for dyads was significantly longer than solos only in the last five trials of the symmetric context (p < 0.01), and no difference was found in other conditions.

Lastly, the total number of trials required to accomplish 40 successful trials was not significantly different between solo and dyadic conditions in the asymmetric context. However, dyads required more trials than solos in the symmetric context (Fig. 2c). One-sample t-tests revealed a significantly larger number of trials in the dyadic than solo performance of the symmetric context (p < 0.001). Furthermore, the total number of trials of dyadic conditions compared to solo conditios of the symmetric context is significantly larger than the asymmetirc condition (p < 0.001). Across all three behavioral metrics, we found that dyads performed better than solo mostly when performing the task in the asymmetric context, whereas solos performed better than dyads mostly in the symmetric context.

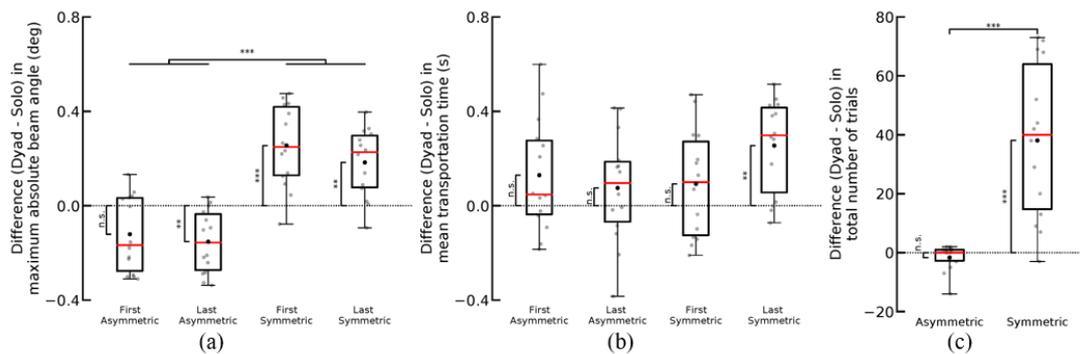

**Figure 2**: **Dyads and solos performance for the asymmetric and symmetric contexts.** **(a).** Between-condition difference in the average maximum absolute beam angle in the first and last 5 successful trials performed in each context. **(b).** Between-condition difference in the average transport times in each context. **(c).** Between-condition difference in the total number of trials performed in each context. Grey dots denote each participant's difference in the metric (solo performance is subtracted from his/her dyadic performance). Asterisks denote statistically significant differences from zero (no difference). *, ** and *** denote p<0.05, 0.01 and 0.001, respectively).

B. *Role specialization emerged during performance in the asymmetric, but not the symmetric context*

Although solos and dyads performed similar movements in our beam transport task, their movement execution was different across task contexts. Figures 3 and 4 show representative time plots of dyadic performance in the asymmetric and symmetric context, respectively. A difference in the agents' position can also be noticed (Fig. 3a and b), i.e., the agents moved differently when performing the task in the asymmetric context.

Specifically, for the asymmetric context the left and right agents applied forces in opposite directions (Fig. 3c). In contrast, these differences disappeared in the symmetric context (Fig. 4a, b and c, d). These observations suggest that participants used different control strategies in the two task contexts to achieve the same goal.

*C.*

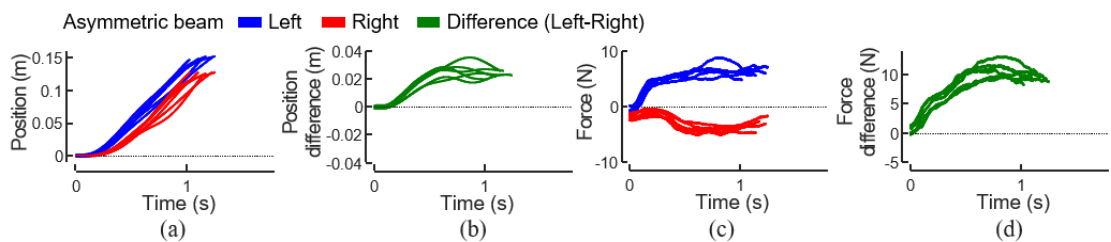

**Figure 3**: **Representative handle position and force trajectories from dyad performance: asymmetric context. (a)**. Right and left handle position. **(b)**. Difference between left- and right-handle positions. **(c)**. Force applied to the left and right handles. **(d)**. Difference between force applied to the left and right handles.

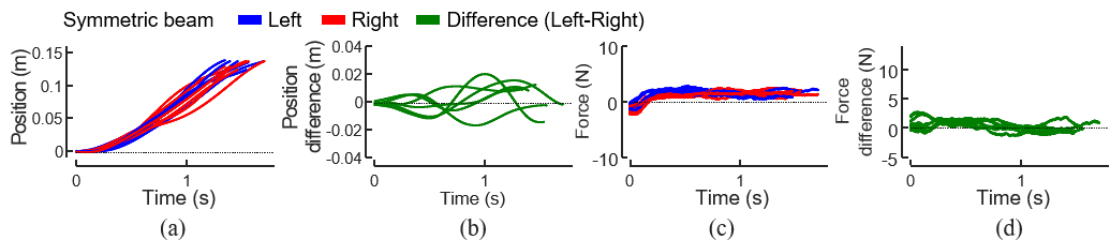

**Figure 4**: **Representative handle position and force trajectories from dyad performance: symmetric context.** Handle position and force data from the same representative dyad shown in Figure 3 are shown in the same format for the symmetric context.

To quantify the spatial coordination between two hands, we subtracted the right handle position from that of the left handle (left minus right) so that a positive value would denote left-handle leading. We show that the left participant's handle within dyads and the left hand of solo participants led the right handle in the asymmetric context, but not in the symmetric context, during the last five successful trials (Fig. 5a). In contrast, during the first five successful trials, small position differences between two hands were found with similar magnitude for both symmetric and asymmetric contexts, in both dyadic and solo conditions.

The above observations were confirmed by statistical analysis. For the dyadic conditions, we found significant main effects of Trial ($F(1,6) = 11.62$, $p = 0.014$) and Context ($F(1,6) = 56.36$, $p < 0.001$). For the solo conditions, we found a significant Trial × Context interaction ($F(1,13) = 24.79$, $p < 0.001$). Post-hoc tests revealed that spatial left-lead was greater during the last than the first five successful trials in the asymmetric context ($p < 0.001$), and that spatial left-lead was greater in the asymmetric than the symmetric context during the last five successful trials ($p = 0.001$). Lastly, one-sample t-tests further confirmed this phenomenon, with only the spatial lead during the last five trials in the asymmetric contexts being significantly different from zero ($p = 0.002$ and $p = 0.003$ for dyadic and solo condition respectively).

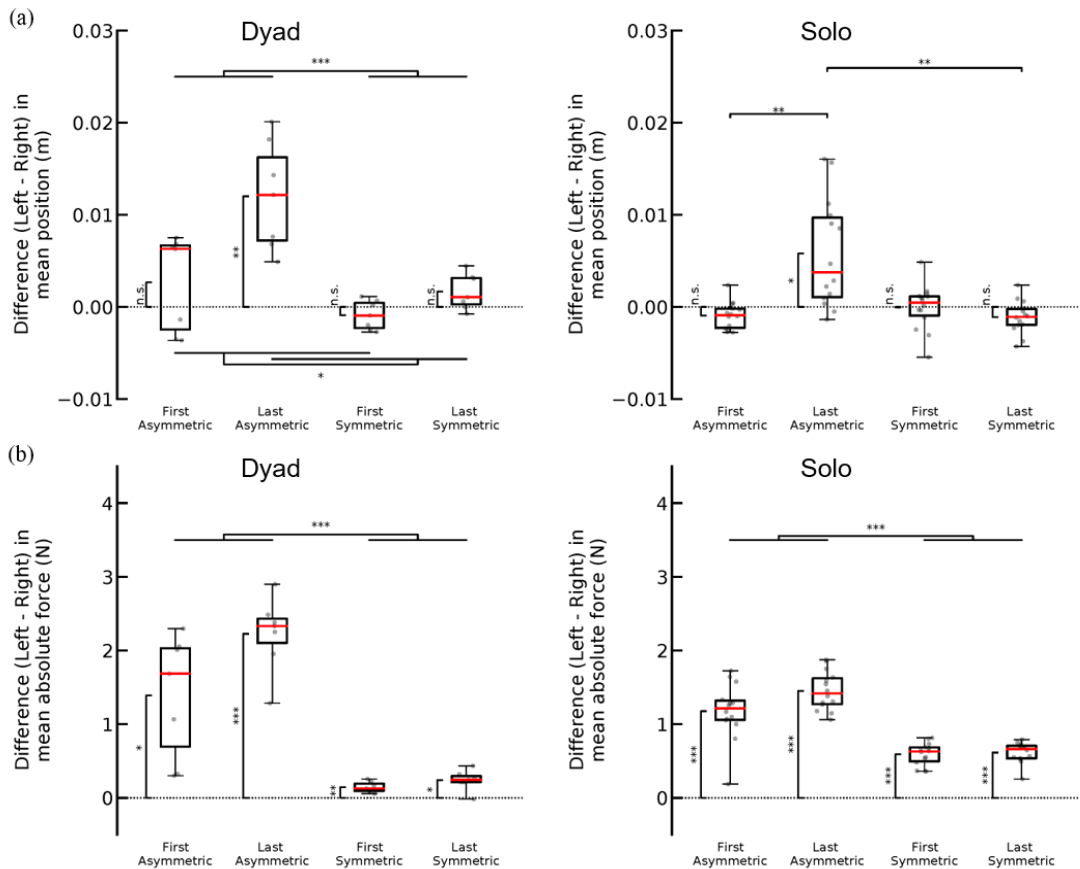

**Figure 5**: **Spatial coordination and measured handle forces: asymmetric and symmetric contexts.** **(a).** Across-condition (Dyadic and Solo) comparisons on differences between left and right mean handle position of the first and last five trials for each task context (asymmetric and symmetric). **(b).** Across-condition (Dyadic and Solo) comparisons on differences between left and right mean absolute handle force of the first and last five trials for each task context (asymmetric and symmetric). Grey dots are data from each dyadic and solo conditions. Asterisks denote statistically significant differences between trials and contexts (*, ** and *** denote p<0.05, 0.01 and 0.001, respectively).

To quantify the relative force contribution from two hands, we quantified the differences in mean absolute force between two hands (left minus right). We found greater force differences in the asymmetric than symmetric contexts during the first and last five trials for both dyadic and solo conditions (Fig. 5b). Specifically, the left participant (or hand) used larger force than the right participant or hand in the asymmetric context. Statistical analysis confirmed the significant main effect of Context for both dyadic and solo conditions (F(1,6) = 127.15, p < 0.001 and F(1,13) = 192.72, p < 0.001, respectively). This suggests a greater left-right force difference in the asymmetric than symmetric condition. We also found a significant main effect of Trial for solo condition (F(1,13) = 5.71, p = 0.033), and one-sample t-tests showed that left-right force difference in all conditions were significantly greater than zero (p < 0.005).

In summary, we found differences in handle position and force when comparing task context across solos and dyads. Specifically, when solos and dyads performed the asymmetric context, one handle consistently led and exerted more force than the other. In contrast, no clear position or force differences were found when solos and dyads performed the symmetric context.

**DISCUSSION**

In the asymmetric context, there was no significant difference in the transport time between dyads and solo participants (Fig. 2b). However, dyads achieved a significantly smaller beam angle than solo participants (Fig. 2a). In contrast, in the symmetric context, performance by solos was characterized by significantly shorter transport times, smaller beam angles, and fewer trials. That is, dyads performed better in the asymmetric context, the only experimental condition where role differences were observed. In fact, in the symmetric context where dyads did not exhibit clear role differences, the beam was less stable and dyads required more transport time and trials than solos. Overall, these results support our hypothesis that spontaneous role specialization enables dyads to outperform solos, rather than better performance arising from two participants cooperating to attain a common goal per se. Therefore, two are not always better than one when performing collaborative tasks.

A. ***Spontaneous role specialization in physical joint actions: dyadic and solo conditions.***

In dyadic physical joint actions, two agents can be assigned to take different roles based on asymmetric constraints imposed by the visuomotor control interfaces. For example, robotic teleoperation tasks may be constrained by different types of physical operations afforded by the robotic slaves[13], or different fields of view[26] each agent may have access to. In these cases, the pre-determined roles specialization is often designed to match the task constraints, e.g., one agent controls the vertical position of the load with a crane whereas the other one controls the horizontal position of the load with a robot[13]. However, many other collaborative tasks may not explicitly impose asymmetric constraints on how each agent can perform the task, and the participating agents are not directly assigned different roles. In these cases, role specialization may emerge spontaneously, and this phenomenon has been quantified in task-specific ways in the literature. Reed and Peskin[15] showed that two agents may specialize on performing the acceleration and deceleration phase of a fast crank rotation task. Groten et al.[16] quantified the force contribution in a haptic collaborative tracking task and found that the contribution was asymmetrical between two agents, but that the asymmetry was not consistent across trials within each pair. Moreover, Melendez-Calderon et al.[18] showed that the torque contributions from two agents in a similar haptic collaborative wrist tracking task can be classified into different categories based on distinct asymmetric patterns. However, these patterns were not consistent within or across dyads. Bosga et al.[17] asked dyads to move a rocking board to track different prescribed oscillatory motions by full body motion and demonstrated that one agent is more likely to temporally lead the dyadic movement.

It is important to point out that the task contexts in the above-mentioned studies were symmetric, that is: the same control input to the task from each agent would have the same effect on the task. It has been argued that the extent to which role specialization can emerge in these tasks can be a result of idiosyncratic differences in individuals' sensorimotor capabilities (e.g., speed, accuracy, strength), which may not require pair-wise co-adaptation to make asymmetrical contributions to the task. For example, differences in reaction time and movement speed could lead dyads to exhibit spatiotemporal asymmetry in joint reaching[27]. Therefore, previous work has not found consistent role specialization in symmetric tasks and has often been quantified on a dyad-to-dyad basis, making it difficult to assess common patterns among dyads. In contrast, our recent work[19] demonstrated that consistent role specialization can emerge gradually to take advantage of the implicit asymmetric task context.

Note that computational simulation in that previous work showed that the asymmetric beam movement task has a large solution space for dyadic coordination.[19] Therefore, it is not necessary for dyads to take different roles to successfully perform the task, although symmetric dyadic coordination (i.e., no spatial leader-follower relation) could result in larger peak tilt. In fact, when the asymmetric task context was performed without haptic feedback, no role specialization was found. Therefore, it is likely that dyads exploited the asymmetry of the mechanical leverage in this task to improve performance through a haptically-mediated co-adaptation process. The experimental results of the asymmetric context in the current study are consistent with the findings of our previous work, showing that the left handle spatially led the right handle (Fig. 3(a)), and the force applied on the handles was uneven (Fig. 3(c)). Furthermore, we also showed that individuals adapted to the asymmetric context using a similar inter-limb coordination strategy. In contrast, when performing the task in the symmetric context, role specialization between two limbs was much weaker in both dyads and solos.

When an individual performs a visuomotor task bimanually, two hands may contribute differently even if the task context is symmetrical. Such role specialization could be attributed to the differences in motor control of dominant and nondominant hands. Because the brain has direct access to the internal models of each hand (e.g., strength and variance), the contribution of each hand to the task can be determined in an optimal fashion[28, 29]. Furthermore, the dynamic dominance framework[30] proposes that control of the dominant and non-dominant hand is predominantly mediated by predictive and reactive control, respectively. Therefore, it is possible that role specialization can directly emerge from allowing each hand to perform in its specialized mode. In one of our previous studies, we showed that the dominant (right) hand can move faster than the non-dominant (left) hand in a symmetric continuous object balance task[20]. Moreover, there is evidence that bimanual coordination may be organized by the dominant hemisphere, as cross-limb coupling is often asymmetric[31, 32, 33]. However, we did not find strong evidence for unequal contributions by the two limbs in when solos performed the symmetric context. Instead, we found a stronger role specialization in the asymmetric context, which was acquired gradually after repeated exposure to the context. It is unlikely that this role specialization was the result of intrinsic differences between the limbs because limb dominance effects do not require practice to be observed. Instead, the emergence of such asymmetric contribution in the solo condition in our results could be explained by a motor learning process in which participants found better coordination strategies to leverage the task asymmetric context.

While both dyads and solo individuals learned to perform the asymmetric context with specialized roles for two hands, the underlying learning process may differ. When learning as an individual, the somatosensory information from both hands can be integrated together with the visual feedback of the object movement to infer the task dynamics, which remains invariant. In contrast, when learning as part of the dyad, an individual can only receive haptic feedback from the consequence of the combination of the task dynamics and the motor actions of the other agent, who may change the motor actions on a trial-to-trial basis. Nevertheless, how such mixed information drives the dyads to adopt the observed coordination pattern remains to be investigated in future studies.

B. *Spontaneous role specialization may facilitate dyadic performance*

Previous studies using tasks with explicitly assigned asymmetric roles revealed that dyads performed better than individuals. For example, when two agents were assigned to control different dimensions of object movement, dyads could complete the tasks faster[13, 34]. In the present study, we found that task performance, i.e., the tilt angle of the beam, was better in dyads than solos for the asymmetric context where consistent dyadic role specialization emerged spontaneously. In contrast, dyadic performance for the symmetric context was worse than solo individuals.

Our results are consistent with the proposition that group benefit may originate from allowing each agent to make contributions that are in line with individual capabilities[6], this phenomenon manifesting as role specialization. Furthermore, our results can be explained by differences in sensorimotor constraints for controlling bimanual versus joint motor actions. Bimanual coordination might be constrained by inter-limb coupling effects such that greater neural resources may be engaged during asymmetric than symmetric bimanual coordination[35, 36, 37]. Therefore, it is conceivable that bimanual performance in our symmetric context might have been cognitively less demanding than the asymmetric context. An opposite scenario occurs when dyads were asked to perform our task in the symmetric context. Here, spatiotemporal synchronization is difficult because sensorimotor delays and noise challenges each agent's ability to match the movement characteristics of the other. In contrast, when asymmetric actions are produced through role specialization, spatiotemporal synchronization is less important and each individual can focus on a unimanual sub-task, leading to lesser cognitive demand for each agent.

Future studies using brain imaging are needed to identify the neural mechanisms involved with haptically-mediated dyadic cooperation.

*C.* **Conclusions**

This study is the first to report the relationship between role specialization and performance in virtual haptically-coupled paired and individual tasks. Dyads outperformed solos but only when clear roles emerged through redundant control of a task in an asymmetric context. In contrast, solos outperformed dyads when role specialization did not occur in a context that required shared control by two agents. We conclude that role specialization, rather than dyadic interaction per se, is a key factor underlying superior performance in dyads. That is, two are not always better than one.

**ACKNOWLEDGMENTS**

This research was supported by a National Science Foundation grant BCS-1827752, the Commissioned Research of NICT, AMED under Grant Number JP21he2202005, Innovative Science and Technology Initiative for Security Grant Number JPJ004596, ATLA, Japan, JST [Moonshot R&D] [Grant Number JPMJMS2034], and JSPS KAKENHI Grant Number JP20K20263.